# On the relation between the width of the flux tube and $T_c^{-1}$ in lattice gauge theories*

M. Caselle[a], F. Gliozzi[a], and S. Vinti[a, b]

[a]Dip. di Fisica Teorica dell'Università di Torino and I.N.F.N., Via P. Giuria 1, I-10125 Torino, Italy

[b]Centro Brasileiro de Pesquisas Fisicas, Rua Dr. Xavier Sigaud 150, 22290 Rio de Janeiro, Brazil

Within the framework of a quantum flux tube model for the interquark potential it is possible to predict that in (2+1) dimensions the space-like string tension must increase with the temperature in the deconfined phase and that the thickness of the flux tube must coincide with the inverse of the deconfinement temperature. Both these predictions are in good agreement with some recent numerical simulations of SU(2) and $Z_2$ gauge models.

## 1. Introduction

It is well known that in finite temperature Lattice Gauge Theories (LGT) the space-like string tension (namely that extracted from Wilson loops orthogonal to the compactified imaginary time direction) is no longer the order parameter of confinement and is, in general, different from zero even in the deconfined phase. Recently, it has been shown in various LGT's, both in (2+1) and (3+1) dimensions that such space-like string tension *increases with the temperature* in the deconfined phase [1–3]. At the same time, it was observed in the SU(2) [2] and $Z_2$ [3] models in (2+1) dimensions that the inverse of the deconfinement temperature $1/T_c$ almost coincides with the thickness $L_c$ of the flux tube joining the quark anti-quark pair in the confined phase.

In this contribution, following our recent paper [3], we want to show that both this results can be understood in the framework of a simple quantum flux tube model. The only assumption that we need in order to obtain these results is that the quantum fluctuations of the flux tube vanish above the deconfinement temperature. Suitable ratios of Wilson loops, which select the contribution of these fluctuations, can be used to test such assumption [3]. We verified that it is correct, using Montecarlo simulations, in the case of the $Z_2$ model [3]. The vanishing of the quantum fluctuations can be considered as a sort of order parameter of a new phase transition, whose temperature coincide with the deconfinement temperature, but which is independent from deconfinement (let us stress once more that space-like Wilson loops cannot experience finite temperature deconfinement). This phase transition is similar to the roughening transition, and should be in the same universality class. The resulting phase diagram strongly resembles that of two dimensional $Z_N$ symmetric spin models.

## 2. Flux tube model

The flux tube model is based on the idea that in the confined phase of a gauge theory the quark-antiquark pair is joined together by a thin, fluctuating flux tube [4]. The simplest version of the model (which should be a good description when heavy quark-antiquark pairs, at large interquark separation $R$, are studied) assumes that the chromo-electric flux is confined inside a tube of small but nonzero thickness $L_c$, that $L_c$ is constant along the tube (neglecting boundary effects near the quarks) and independent of the interquark distance. An immediate consequence of this picture is linear confinement: the potential $V(R)$ rises linearly according to the law $V(R) = \sigma R$. A second important consequence is that the string tension $\sigma$ and the effective cross-section of the flux tube, $A_t$ are related by the

---

law [5]

$$\sigma = \frac{c_t}{A_t} \qquad (1)$$

with $c_t$ a suitable constant. Eq (1) can be tested, for instance, by measuring the string tension on asymmetric lattices, with the size (say $L_s$) in the direction orthogonal to the Wilson loop, smaller than the others. If $L_s > L_c$ no effect on the string tension is expected. When, on the contrary, $L_s < L_c$, then the flux tube is squeezed. If we assume for simplicity that the flux density is uniform inside the flux tube, namely then $A_t \propto L_c^{(d-2)}$ (where $(d-2)$ is the number of transverse dimensions) and eq.(1) suggests, for these values of $L_s$, a linear rising of the string tension according to the law:

$$\sigma(L_s) = \frac{L_c}{L_s}\sigma(\infty) \quad , \qquad (2)$$

where $\sigma(\infty)$ denotes the string tension in the uncompressed situation, namely for $L_s \gg L_c$ (in the following we will denote $\sigma(\infty)$ with $\sigma$ for brevity). This law gives us a powerful tool to probe the interior of the flux tube and to test the range in which the uniform density approximation is valid. For instance eq.(2) is well satisfied in the range $L_c/2 < L_s < L_c$ in the case of the the $(2+1)$ dimensional $SU(2)$ [2] and $Z_2$ [3] models .

An obvious remark at this point is that the situation that we have described is exactly the same that one finds in finite temperature LGT's if Wilson loops orthogonal to the (imaginary) time direction are studied. What is remarkable, is that in both the $Z_2$ and the $SU(2)$ model the resulting value of $L_c$ turns out to be very similar to the inverse of the deconfinement temperature [2,3]. In order to understand this result we must address the question of the quantum fluctuations of the flux tube.

It is by now generally accepted that these quantum fluctuations can be effectively described by a massless two-dimensional free field theory, where the free field $h(x_1, x_2)$ describes the displacement from its equilibrium position of the surface bordered by the Wilson loop. This idea traces back to the seminal work of Lüscher Symanzik and Weisz [6,7] and has been discussed from then in several papers (see for instance ref [8], and references therein).

Let us mention two interesting consequences of quantum fluctuations.

First, by integrating over $h(x_1, x_2)$, the Wilson loop expectation value $W(L_1, L_2)$ may be written in the form

$$W(L_1, L_2) = e^{-\sigma L_1 L_2 + p(L_1 + L_2)} Z_G \quad , \qquad (3)$$

where the contribution $Z_G$ of the quantum fluctuations can be expanded as follows

$$Z_G(L_1, L_2) = \tilde{C}(L_2)^{-\frac{1}{4}} q^{-\frac{1}{48}} \sqrt{1 + q + 2q^2 \dots} \quad (4)$$

where $\tilde{C}$ is an undetermined constant, $q = exp(-2\pi\frac{L_1}{L_2})$ and we have assumed , without loss of generality, $L_1 \geq L_2$. The quantum correction $Z_G$ can be precisely measured using Montecarlo simulations.

The second result emerges if one studies finite temperature LGT's. Looking at the quantum fluctuations of the surface bordered by two Polyakov loops it is possible to predict the ratio between the deconfinement temperature and the square root of the zero-temperature string tension $\sigma$ [9]:

$$\frac{T_c}{\sqrt{\sigma}} = \sqrt{\frac{3}{\pi(d-2)}} \qquad (5)$$

where $(d-2)$ is the number of transverse dimensions of the flux tube.

Let us make at this point our main assumption. We assume that the *quantum corrections $Z_G$ should disappear* when the flux tube fills the whole lattice, namely when $T > T_c$ or $L_s < L_c$, depending on the geometry we are interested in. Intuitively this is equivalent to assume some sort of selfavoiding behaviour of the flux tube, since in this case , when the flux tube fills the whole lattice there is no more space left for it to fluctuate. This assumption is in good agreement with recent Montecarlo simulations of the $(2+1)$ $Z_2$ model [3]. This vanishing of quantum fluctuations can be described in a more rigorous way by noticing that the compactification in one lattice direction (say, $L_s$) naturally induces a compactification of the field $h(x_1, x_2)$ on a circle of radius $R = \frac{L_s}{2\pi}$. The quantum field theory of a two-dimensional bosonic field compactified on a circle is by now rather well understood. In particular the spectrum of states and the partition function are exactly known (at least for the so called "rational models" for which $R^2$ is a rational number). It is thus possible to follow the behaviour of the quantum corrections as a function of $R$: $Z_G = Z_G(R)$. What is interesting is that there are precisely four values of $R$ (but only two of them are independent, the others being related by duality) for which the contribution of such quantum corrections vanishes (they correspond to the so called "topological field theories", see ref. [8] for notations and bibliography on this subject). Following the arguments of ref. [8,10], we can predict that the value (let us call it $L_0 = 2\pi R_0$) of the lattice size which corresponds to a zero-contribution point must be related to the (zero temperature) string tension as follows

$$L_0 = \sqrt{\frac{\pi}{3\sigma}} \quad . \tag{6}$$

Following our assumption we can thus say that *moving toward higher temperatures corresponds in the flux tube effective model to a flow toward one of these zero-contribution points* and that $L_0$ must coincide with $L_c$, the flux tube thickness. By comparing eq.s (6) and (5) we then see that in (2+1) dimension the flux tube thickness $L_c$ exactly coincides with the inverse deconfinement temperature.

The fact that at $L_s = L_c$ the quantum fluctuations of the flux tube vanish, strongly resembles what happens at the roughening point, and suggests that at $L_s = L_c$ a new phase transition occurs, in the same universality class of the roughening one. This conjecture is supported by the following argument. When we study the expectation value of a Wilson loop, we are actually breaking the translational invariance in the direction orthogonal to the Wilson loop. Since this is a continuous symmetry, due to the Mermin-Wagner theorem, it cannot be spontaneously broken in two dimensions (the plane where the Wilson loop is defined). The roughening transition is indeed the point where this translational symmetry is recovered [7]. The resulting phase diagram is similar to that of the two dimensional XY model, the roughening point and the rough phase being respectively the Kosterlitz-Thouless point and the critical phase of the XY model. When we compactify the direction orthogonal to the Wilson loop, and choose a finite value of its size $L_s$, we are actually substituting the (continuous) translational group with a (finite) $Z_{f(L_s)}$ group, where $f(L_s)$ is a suitable, finite, function of $L_s$. Then the Mermin-Wagner theorem does not apply anymore. Indeed it is well known [11] that the $Z_N$ models (for $N > 4$) have two Kosterlitz-Thouless points, related by duality, a critical phase in between and two non-critical phases at high and low temperatures. This is exactly our situation, the critical region being the one in which quantum fluctuations exist (and, more generally, where the methods of 2d QFT apply) and the high temperature non-critical region being equivalent to the "squeezed flux tube" phase (or equivalently the $T > T_c$ phase of space-like Wilson loops) described in the present contribution.